\input{aipcheck}

\documentclass[
    ,final            
  ]
  {aipproc}

\layoutstyle{6x9}

\begin{document}

\title{Formation of quark phases in compact stars and SN explosion}


\author{A. Drago}{
  address={Dipartimento di Fisica, Univ. Ferrara and INFN Ferrara}
}

\author{G. Pagliara}{address={Institut f\"ur Theoretische Physik, J.W. Goethe Universit\"at}}

\author{G. Pagliaroli}{
  address={Dipartimento di Fisica, Univ. L'Aquila and INFN Gran Sasso}} 

\author{F.L. Villante}{address={Dipartimento di Fisica, Univ. L'Aquila and INFN Gran Sasso}}
\author{F. Vissani}{address={INFN Gran Sasso, Theory Group}}

\begin{abstract}
We describe possible scenarios of quark deconfinement in compact stars and
we analyze their astrophysical implications.
The quark deconfinement process can proceed rapidly, as a strong
deflagration, releasing a huge amount of energy in a short time
and generating an extra neutrino burst.
If energy is transferred efficiently to the surface, like e.g. in the presence of
convective instabilities, this burst could contribute to revitalize a
partially failed SN explosion.
We discuss how the neutrino observations from SN1987A would fit in this
scenario. Finally, we focus on
the fate of massive and rapidly rotating progenitors, discussing
possible time separations between the moment of the core collapse and
the moment of quark deconfinement.
 This mechanism can be at the basis of the
interpretation of gamma ray bursts in which lines associated with heavy
elements are present in the spectrum.
\end{abstract}

\maketitle


\section{Introduction}
The mechanism of core collapse Supernova (SN) explosions has not yet
been completely clarified, although relevant progresses have been made
in the last years. It is possible that, while SNe with a relatively light progenitor
can explode via the standard mechanism,
new physical ingredients are needed to explain SNe with progenitor
masses larger than $\sim 20 M_\odot$.

In the last years a huge amount of papers has been published, discussing the
existence of deconfined quarks inside a compact star. Quark deconfinement inside
a compact star is an exothermic process releasing in general a large amount of energy.
Thus, it is interesting to explore its astrophysical implications.
In this contibution we will first describe possible scenarios of quark deconfinement
in compact stars. We will then focus on the impact of this process
on SN explosion and on the connection between SNe and Gamma Ray Bursts
(GRBs).

\section{When and How to deconfine inside a compact star?}
In the literature several scenarios of quark deconfinement have been discussed,
and they can roughly be grouped in two categories, depending on when
the deconfinement transition takes place:
\begin{itemize}
\item
Quark Deconfinement Before Deleptonization (QDBD) of the
protoneutron star;
\item
Quark Deconfinement during or After the Deleptonization (QDAD).
\end{itemize}
Concerning the first possibility, QDBD,
the main idea is that during the collapse of the homologous core the
first critical density is reached, separating the pure hadronic phase from the mixed
phase of quarks and hadrons\footnote{
An interesting variation of this possibility has recently been
discussed
\cite{frankfurt}, in which the mixed phase is mainly produced 
not at the moment of
the core bounce, but during the fallback of the material following a failed SN explosion.} \cite{Gentile:1993ma,Drago:1997tn,Nakazato:2008su}.
Since in the mixed phase the adiabatic index is very low,
the collapse continues rapidly through the mixed phase till the central density
reaches the second critical density separating mixed phase and pure quark matter.
At this point, the adiabatic index becomes large again and the collapse halts. A shock wave is then
produced. One feature of this mechanism is that
it requires a particularly soft
Equation of State (EoS), since the formation of a mixed phase of quarks and
hadrons has to take place at the relatively low densities reached at the moment of
core bounce (or immediately after, during the fallback but anyway before deleptonization
\cite {frankfurt}).
Since the densities reached at the moment of the bounce are only moderately dependent
on the mass of the progenitor, this mechanism is rather ``universal'', affecting most
of the SNe, although its effect on the explosion can still depend on the mass of the progenitor.

The second possibility, QDAD, is that quark deconfinement takes places only after an at least
partial deleptonization \cite{Pons:2001ar,Aguilera:2002dh}. It is well known, in fact, that when the pressure due to leptons
decreases, the central baryonic density increases and therefore the deconfinement
process becomes easier\footnote{
It is also important to recall that a important role
is played by strangeness: is it much more easy to deconfine into strange quark matter than
into purely up and down quarks. On the other hand one needs to clarify how the strange
quarks are produced, if they are not already present in the hadronic phase.}.
Clearly, a temporal separation can exist in this second scenario between the moment
of core collapse and the moment in which quarks deconfine.

Another important question concerns the way in which the deconfinement transition takes place,
either as a smooth transition, in which no surface tension is present and no metastability
(of the hadronic phase) can exist or, instead, as a first order transition in which
a surface tension exists at the interface between hadrons and quarks.

The first possibility, i.e. no metastability,
was analyzed in a QDBD scenario in \cite{Gentile:1993ma,Drago:1997tn,Nakazato:2008su,frankfurt}
where the time scale of the formation of the mixed phase depends only on the 
velocity of compression. Here a very important role is played by
the mechanical shock wave which forms at the interface between the mixed phase and the pure quark phase,
which could help the SN to explode as shown in \cite{frankfurt} for low- and 
intermediate-mass progenitor stars. 
Within the
QDAD scenario of Ref.~\cite{Pons:2001ar} deconfinement is also a gradual process,
driven in this case by deleptonization. The total thermal energy emitted in neutrinos increases, but the
neutrino luminosity is almost unchanged and in particular no new neutrino burst is obtained
in association with deconfinement.

The existence of surface tension between quarks and hadrons opens the
possibility that hadronic matter remains metastable for a short or
long time period. The duration of the metastable phase depends mainly
on: i) the numerical value $\sigma$ of the surface tension; ii) the rate by
which the parameter controlling the degree of metastability (for
instance the density of the system, or the leptonic content) changes
with time. A not too small of $\sigma$ and a slow change in the
density can allow the system to remain metastable even for a very
long time \cite{Berezhiani:2002ks}. On the other hand, even a very short period of
metastability (order of seconds)
can be extremely important, because it allows the system to evolve on
time scale dictated by the velocity of the burning. This point is particularly
relevant because we will show that the burning of hadrons into quarks
inside a compact star is a strong deflagration and not a detonation.

{\it In the following we will concentrate on the QDAD scenario in presence
of surface tension which implies the formation of quarks
via drops nucleation.}

\begin{figure}
  \includegraphics[height=.2\textheight]{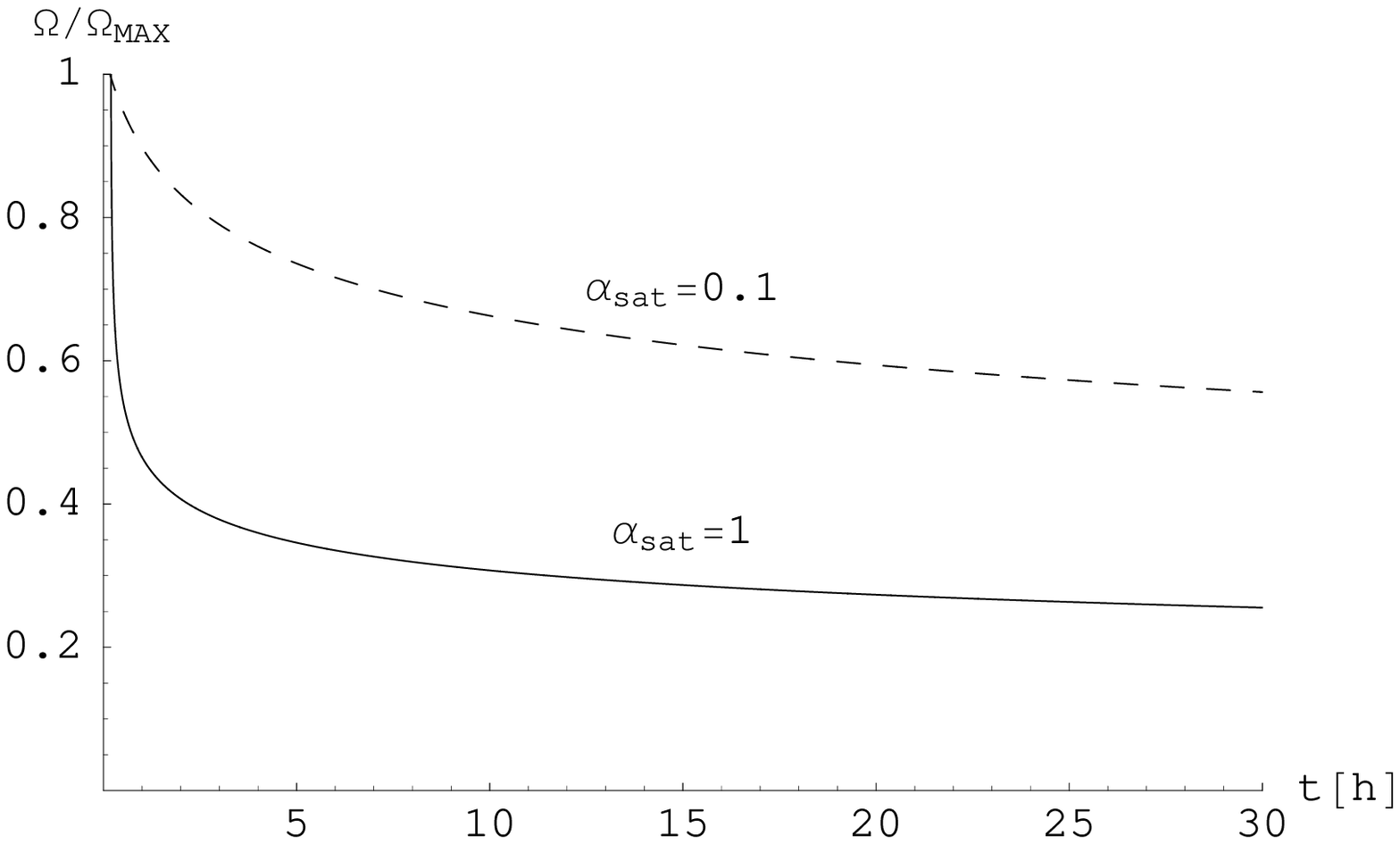}
  \includegraphics[height=.2\textheight]{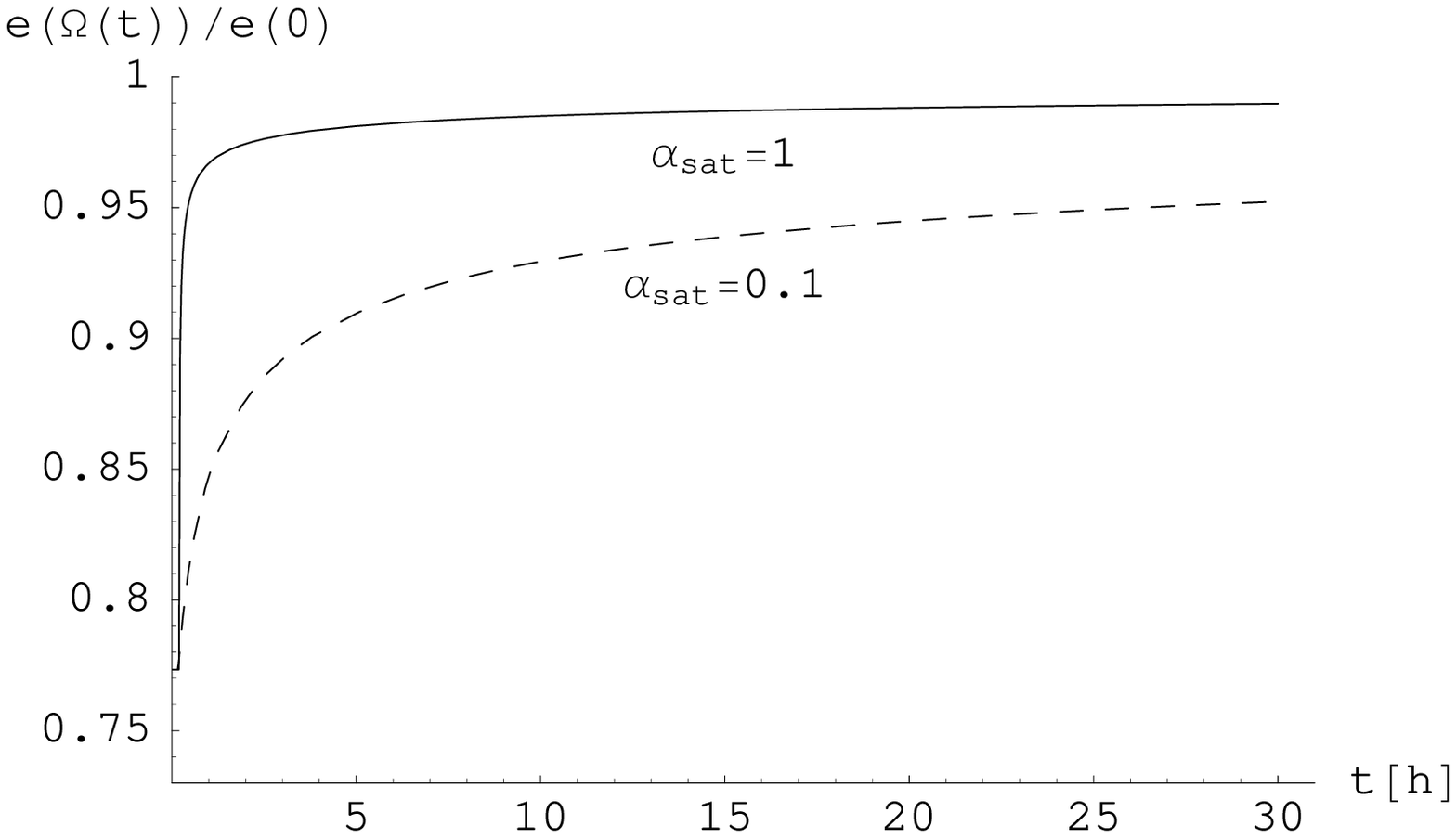}
  \caption{Left panel: Angular velocity (in units of the Keplerian velocity)
  as a function of time. Here the slow down is due to r-modes. Two different values of the
  saturation amplitude are considered \cite{Andersson:2000mf}.
  Rigth panel: Change of the central energy density as a function of time (the rotating configurations
are constructed by using the RNS code \cite{Stergioulas:1994ea}).  }
  \label{rotazione}
\end{figure}

An important role in the scenarios outlined above is played by the
rotation of the star. A rapid rotation produces a very slow
variation of the central density of the compact star. The density will
increase due to the reduction of the angular momentum or
due to the fallback associated with a failed explosion.
As a consequence, the high densities necessary for the deconfinement
are reached after a relatively long time, even of the order of hours.
In Fig. \ref{rotazione} we show an example of the relation between
angular momentum and central density of the compact star.
Here the slow down is due to r-mode instabilities, 
but a strong magnetic field could play a similar role
in rapidly reducing the angular velocity of the star.

\section{Burning of hadrons into quarks inside a compact star}

Let us recall a few results concerning the hydrodynamics of the
combustion of hadrons into quarks in compact stars. Due to the large
density of the system and to the relatively low temperatures involved\footnote{
The estimated average temperature of the quark phase after the
burning is typically below 50 MeV.} the process does not start as a
detonation, but as a strong deflagration. This implies that the
velocity of the front is subsonic and that no shock wave is associated
with the burning. On the other hand a strong deflagration in the
presence of gravity is characterized by an unstable front, where
wrinkles can form. Since a deflagrative front proceeds due to the
transmission of some ``fuel'' across the surface (in this case the
fuel can be strange quarks) and/or the exchange of heat from the
burned zone to the unburned zone, the formation of wrinkles can
significantly increase the burning velocity, since it increases the
area of the surface separating burned and unburned material.

In Ref.~\cite{Drago:2005yj} the velocity of the front has been estimated
taking into account the hydrodynamical instabilities of the burning
front. The net effect of the wrinkles is to increase the burning
velocity up to 10$^3$--10$^4$ km/s, so that the central region of the
compact star can transform into quarks or into a mixed phase on a
time scale of the order of 10$^{-3}$--10$^{-2}$ s. These velocities,
although very large, are clearly subsonic (the velocity of sound in a
compact star is of the order of 10$^5$ km/s) and therefore the
deflagration does not transform into a detonation due to the
hydrodynamical instabilities (at variance with what happens in SNIa).
It is also important to note that the velocity slows down while
the combustion front approaches the surface of the star.

Finally, we remark that convective instabilities can
develop close to the center of the star, due to the different
EoS describing a newly formed bubble of quarks with respect to the
surrounding hadronic material. This type
of convection (not based on a thermal gradient) can develop, however, only in a
relatively narrow range of parameter values, typically the ones which
lead to the formation of a quark star and not of a hybrid star.

\section{Heat transport inside a newly formed hybrid star}
\begin{figure}
  \includegraphics[height=.3\textheight,width=.43\textwidth]{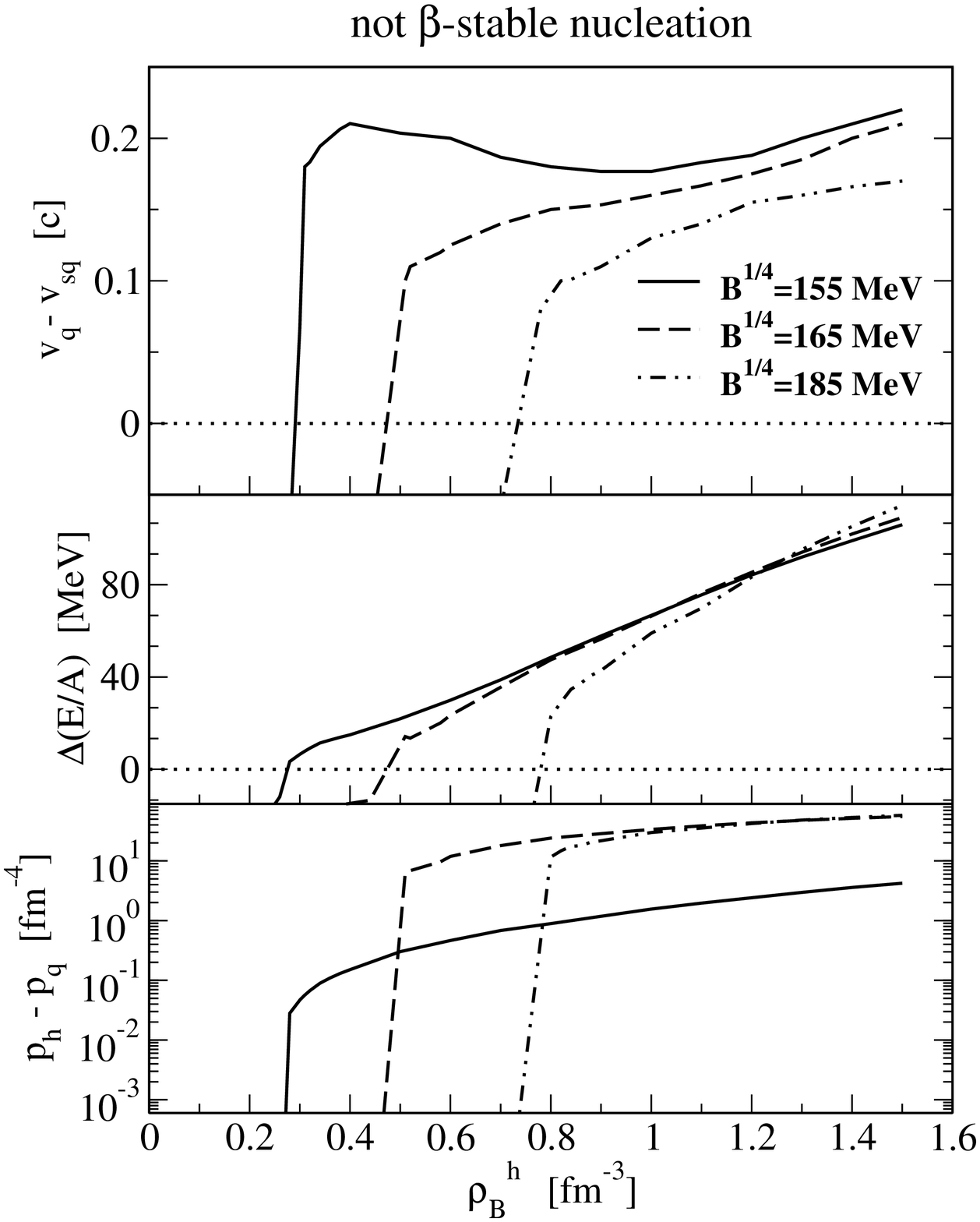}
  \includegraphics[height=0.3\textheight,width=.57\textwidth]{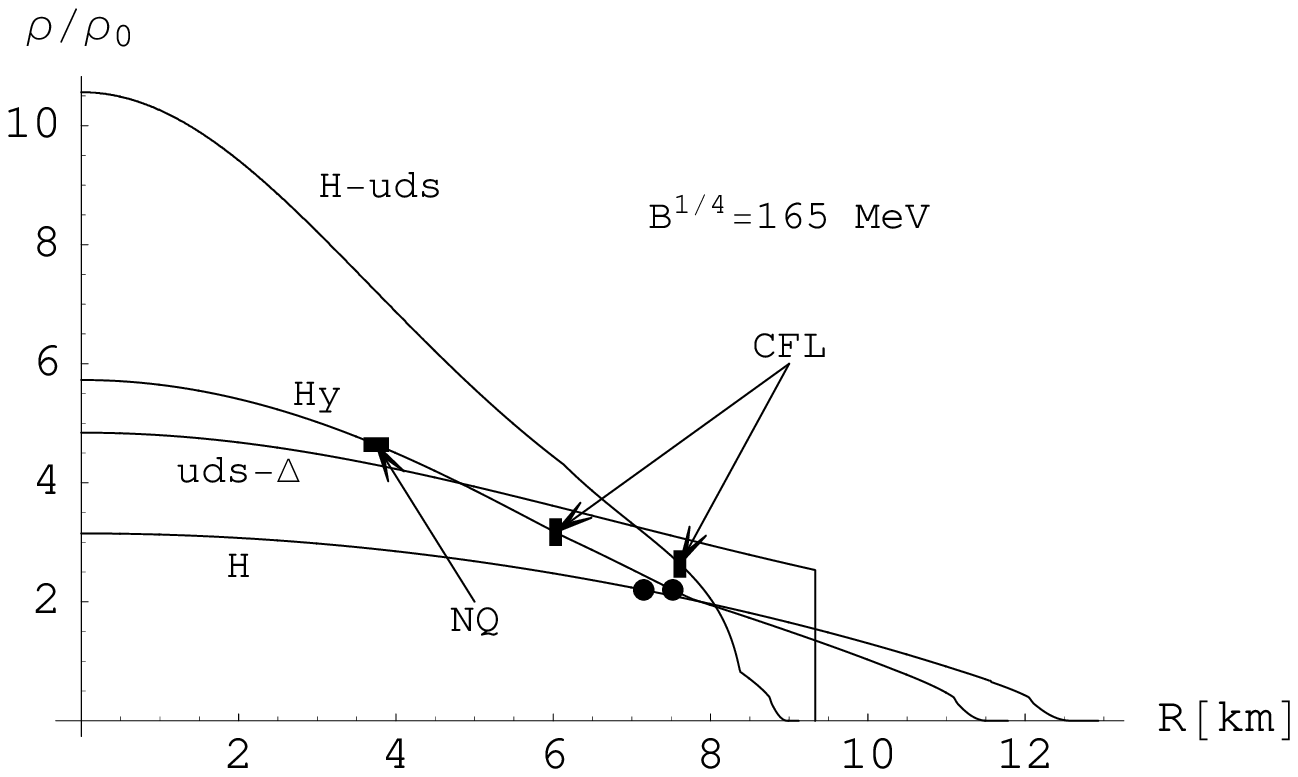}
  \source{Ref. \cite{Drago:2005yj}}
  \caption{Left panel: Burning of hadrons into quarks. Right panel: Baryon density profile of compact stars in various models.
  See text for details.}
  \label{heger}
\end{figure}

The combustion of hadrons into quarks releases a substantial amount of energy \cite{Bombaci:2000cv,Drago:2004vu},
$E_{\rm dec}\sim 10^{53}$ erg,
in the central region of the star and in a time scale of the order of
$\tau_{\rm dec} \sim10^{-3}-10^{-2}$ s.
To consider specific models, we can see
in the left panel (middle plot) of the Fig.~\ref{heger} the energy difference between the
combusted and the uncombusted phases as a function of the baryonic density
and for three values of the MIT bag model parameter $B$.
In the right panel we also show the baryonic density profiles of different
compact stars.
For definiteness, we select the specific case of a hybrid star
corresponding to a quark phase described by the MIT bag model with $B^{1/4}=165$ MeV.
We consider normal not-superconducting ("H-uds" line) as a reference model.

The combustion occurs in the region where $(\Delta E/A)\ge 0$ which corresponds to
density $\rho_{\rm B} \ge 2-3 \, \rho_0$ and extends out to a radius of $\sim 8$ km
(see the kink in the "H-uds" curve in Fig.~\ref{heger}).
One can estimate that most of the energy is released in radial shells at
a distance of $D\sim 5$ km from the surface.
This region is heated to temperatures which can be as large as $T\sim 30$ MeV, much larger
than the temperature of the external hadronic region.

The produced energy is transported to the surface by neutrinos that reach thermal
equilibrium on weak interaction time scales which are much
shorter than $\tau_{\rm dec}$.
We indicate with $\gamma=E_\nu/E_{\rm dec}$ the fraction of the total energy
which is stored in neutrinos at a given moment (before the star cools down significantly).
This quantity depends on temperature and
chemical potentials and it is typically of the order of few percent.

Neutrinos diffuse outwards with a time scale $\tau_{\rm diff}\sim D^2 / \lambda \sim 10{\rm s} \;(D/5\;{\rm km})^2/(\lambda/1{\rm cm})$ where
$\lambda$ is the neutrino mean free path. After the combustion, we thus expect an extra
neutrino burst with a risetime $\tau_{\rm rise}\sim \tau_{\rm diff}$
and a peak luminosity equal to about $L_\nu \sim E_\nu / \tau_{\rm diff} \sim 10^{52}{\rm erg\;s^{-1}} \; \gamma \, (\lambda/ 1{\rm cm})$.
Due to neutrino emission the star will cool (and the neutrino luminosity will decrease)
with a time scale $\tau_{\rm cool}\sim \tau_{\rm diff}/\gamma$.

 It should be noted that the burning of hadrons into quarks produces a large temperature gradient
between the combusted and the uncombusted region which can lead to a convective instability. The existence of a
convective motion which mixes the upper layers of the burned zone with the unburned region is an interesting
possibility which will be explored quantitatively in a future paper.
Convection is a very efficient heat transport mechanism. If the convective layer is sufficiently
extended, the energy released by combustion is transported
to the surface much faster than in the non-convective case.
Moreover, convection brings hot material to the surface which can be heated to a
temperature of the order of few MeV. As a consequence, the neutrino luminosity
is expected to increase on the time scale of convective heat transfer,
reaching a much larger peak value than in the normal diffusive case described above.
If a luminosity equal to about $10^{53} \,{\rm erg}\;{\rm s^{-1}}$ is achieved,
the extra $\nu_e\overline{\nu}_e$-neutrino burst could revitalize the SN explosion.

\section{Comparing with SN1987A observations}
The only SN neutrinos detected till now are those from SN1987A.
Indeed, on February 23, 1987, at
2 h 53 m (UT) LSD detector observed 5 events \cite{lsd};
at 7 h 36 m (UT) IMB, Kamiokande-II and Baksan \cite{resto}
detectors observed 8, 11 and 5 events respectively.
The progenitor was a blue supergiant with estimated mass of $\sim$ 20 $M_\odot$.

The observations of
Kamiokande-II, IMB and Baksan can be explained
very well within the standard scenario for core collapse SNe,
assuming that the events are due to $\bar\nu_e p\to n e^+$.
The observations are consistent with the presence of an initial,
high luminosity phase of neutrino emission,
followed by a thermal phase due to the cooling of the
newborn neutron star \cite{ll}.
Such an  initial and luminous phase
is expected; indeed, it
should trigger the subsequent explosion of the star.
The standard scenario for core collapse SNe
does not predict the existence of multiple pulses
of neutrino emission and thus cannot accomodate LSD data.

Non-standard scenarios with multiple phases of neutrino emission
have been proposed \cite{deruj,imsh}. An interesting
possibility is that the first burst is due to a very intense
neutronization phase by $e^- p\to n \nu_e$;
it was noted in \cite{imsh} that electron neutrinos with an
energy of $30-40$ MeV can be more easily seen in LSD
detector than in the other detectors.
In the astrophysical scenario of \cite{imsh},
the rapid rotation of the collapsing core leads to
a delay between the first and the second
burst.
However, the nature of the second burst is not discussed
in \cite{imsh} and one could doubt whether the beginning
of the second burst includes a phase of initial luminosity.

Here, we consider the possibility that Kamiokande-II, Baksan and IMB
observations are due to the burning of hadrons into quarks.
More specifically, the process of quark deconfinement can provide
the necessary amount of energy in neutrinos, and the occurrence
of convective processes can release a part of this energy
in a short time scale. The intense neutrino luminosity obtained
in this way could not only meet the observations, but also
play a key role for the explosion of the star.

This becomes even more interesting in the presence of a rapid rotation.
The sequence of events, in this case, could be the following:
(1)~an initial intense phase of neutronization accounts
for LSD observations as in \cite{imsh};
(2)~the rapid rotation of the core leads to
the formation of a metastable neutron star, that looses
its angular momentum in a time scale of several hours;
(3)~the central density of the metastable star becomes large enough that
deconfinement can take place.
Again, the rapid release of energy at the beginning of the last stage
could be sufficient to lead to the explosion of the star.

\section{Astrophysical scenarios of explosions driven by
quark deconfinement}

\begin{figure}
  \includegraphics[height=.45\textheight,angle=-90]{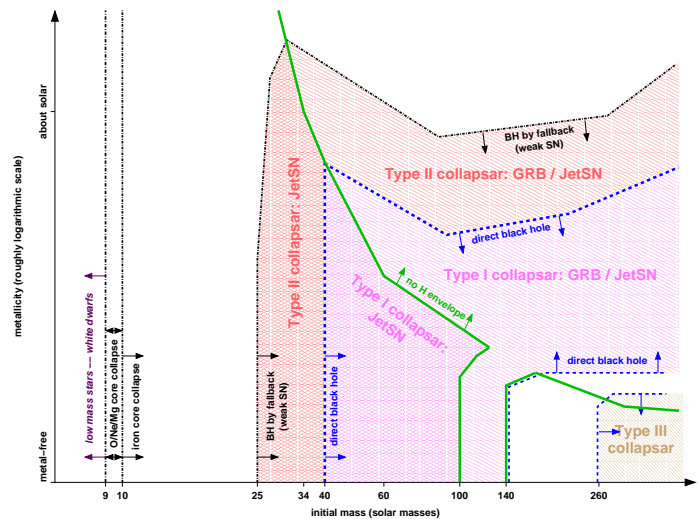}
  \source{Ref. \cite{Heger:2002by}}
  \caption{Fate of massive stars as a function of the initial mass and of the metallicity.}
  \label{Fryer}
\end{figure}

In this section we discuss how quark deconfinement can affect the
standard scenario of the fate of massive stars, as e.g. outlined in
the classical paper of Heger et al.~\cite{Heger:2002by}. In
Fig.~\ref{Fryer} there is a line separating the stars which end up
their life as neutron stars and the ones which produce a black hole by
the fallback of the material not ejected due to a (partially) failed
SN explosion.  This line is particularly relevant to our discussion
because it is located inside the region of large progenitor masses
where the standard mechanism has difficulties in exploding SNe: the
explosion needs to be revitalized by some new injection of energy. We
suggest that the mechanism providing the new energy is based on quark
deconfinement
\cite{Fryer}.

In the scenario described in Fig.~\ref{Fryer} the rotation of the star is crucial
and it plays a double role: it
slows down the fallback and it allows the formation of a jet
which, for high metallicity stars, is at the basis of the
mechanism of formation of a GRB. Let us clarify which scenarios
can open when quark deconfinement takes place inside a rapidly
rotating star:
\begin{itemize}
\item
if the rotation velocity is
large but not extreme, the main effect is the reduction of the
fallback rate. From one side, this reduces the luminosity of the
neutrino burst typically associated with the fallback of the material
on the protoneutron star (therefore making the SN more difficult to
explode); on the other hand, preventing the immediate formation of a black hole,
the rotation allows the neutrino produced by quark
deconfinement to play a important role, since they can have the time
to be produced and to push the star envelope halting the collapse;
\item
if the rotational velocity of the central region is extreme,
the duration of the collapse of the core {\it{before}} the formation of the
protoneutron star can be significantly extended. The peak in neutrino
luminosity associated with electron capture can be temporally
separated from the peak due to quark deconfinement. In principle these
two peaks can be separated by time intervals ranging from minutes to
hours to days or more, depending on the rotational velocity and on the
rate of the fallback \cite{Berezhiani:2002ks}.
\end{itemize}

Are there any observational hints of the scenario outlined above?
We are clearly looking for events in which a first (partially
failed?) SN explosion is followed by a new neutrino outburst 
\footnote{Interestingly, a neutrino burst associated with quark deconfinement
and temporally separated from the first neutronization burst 
was recently discussed in \cite{frankfurt} in a non-rotating scenario.}. A possibile
example is the one already discussed, i.e. the neutrino signals
possibly associated with SN1987A. There, the first partially failed
SN is due to the neutrinos produced at the moment of the prolonged
neutronization which takes place (in the scenario of Ref.~\cite{imsh})
inside the rapidly rotating disk collapsing at the core of the
progenitor star. The moderate fallback due to the rapid rotation
allows the central density of the protoneutron star to gradually
increase till a critical density is reached and a new burst of
neutrinos is generated due to deconfinement. The time interval between
the first and the second burst is regulated by the slowdown of the
protoneutron star, due e.g. to r-mode instabilities or to the presence of a strong
magnetic field.

A second example is provided by GRBs. There have been in the past
several hints of a presence of iron lines in GRBs (either in
absorption or in emission \cite{Amati:2000ad}). Since iron can only be produced at the
moment of a SN explosion, then the SN (even if marginally failed) has
to precede the GRB by a time interval which can be estimated from
data. In our scenario such a delay between SN and GRB can again be due to
the time separation between a (partially failed) SN explosion and the
gamma burst produced by the quark deconfinement neutrinos. A recent
observation based on a Nickel line \cite{Margutti:2007ew} indicates
a time interval of the order of one hour. Other hints of GRBs in which
the SN explosion precedes the GRB are: 
\begin{itemize}

\item  the possible association of GRB with SNIIn \cite{Woosley:2006fn}.
In the standard scenario of GRBs this association should not be
possible, since the GRB would be absorbed by the external shells of
the progenitor; in our case, the time separation between the two
events can help the GRB to emerge from the thick environment;

\item the
evidence of GRBs where no SN is observed \cite{Fynbo:2006mw}. In our scenario, a time
delay between the SN and the GRB of several days or longer would not
allow to observe any signal associated with the SN, which would be too
weak when the GRB signal fades away \cite{Dermer:2007tj}.

\end{itemize}

Finally, a well known problem is the difficulty to produce the heaviest elements through r-processes
taking place at the moment of a SN explosion.
It is tempting to imagine that the ejection of a fraction of
hot and neutron rich material {\it{after}} the first explosion can
allow the production of the heaviest elements via revitalized r-processes
\cite{Jaikumar:2006qx,frankfurt}.

A preliminary version of this work was presented at the Workshop
"The complex physics of compact stars", Ladek Zdroj - Poland, February 2008.

\bibliographystyle{aipproc}

\begin{thebibliography}{9}


\bibitem{Gentile:1993ma}
  N.~A.~Gentile, M.~B.~Aufderheide, G.~J.~Mathews, F.~D.~Swesty and G.~M.~Fuller,
  Astrophys.\ J.\  {\bf 414} (1993) 701.

\bibitem{Drago:1997tn}
  A.~Drago and U.~Tambini,
  J.\ Phys.\ G {\bf 25} (1999) 971

\bibitem{Nakazato:2008su}
  K.~Nakazato, K.~Sumiyoshi and S.~Yamada,
  Phys.\ Rev.\  D {\bf 77} (2008) 103006

\bibitem{frankfurt}
I.~Sagert, M.~Hempel, G.~Pagliara, J.~Schaffner-Bielich,
T.~Fischer, A.~Mezzacappa, F.-K.~Thielemann 
and  M.~Liebend\"orfer, Phys.Rev.Lett., submitted.

\bibitem{Pons:2001ar}
  J.~A.~Pons, A.~W.~Steiner, M.~Prakash and J.~M.~Lattimer,
  Phys.\ Rev.\ Lett.\  {\bf 86} (2001) 5223


  \bibitem{Aguilera:2002dh}
  D.~N.~Aguilera, D.~Blaschke and H.~Grigorian,
  Astron.\ Astrophys.\  {\bf 416} (2004) 991

\bibitem{Berezhiani:2002ks}
  Z.~Berezhiani, I.~Bombaci, A.~Drago, F.~Frontera and A.~Lavagno,
  Astrophys.\ J.\  {\bf 586} (2003) 1250


\bibitem{Andersson:2000mf}
  N.~Andersson and K.~D.~Kokkotas,
  Int.\ J.\ Mod.\ Phys.\  D {\bf 10} (2001) 381




\bibitem{Stergioulas:1994ea} 
N.~Stergioulas and J.L.~Friedman,
Astrophys. J. {\bf 444} (1995) 306.


\bibitem{Drago:2005yj}
  A.~Drago, A.~Lavagno and I.~Parenti,
  Astrophys.\ J.\  {\bf 659} (2007) 1519

\bibitem{Bombaci:2000cv}
  I.~Bombaci and B.~Datta,
  Astrophys.\ J.\  {\bf 530} (2000) L69

\bibitem{Drago:2004vu}
  A.~Drago, A.~Lavagno and G.~Pagliara,
  Phys.\ Rev.\  D {\bf 69} (2004) 057505


\bibitem{lsd}
M.~Aglietta {\it et al.},
  Europhys.\ Lett.\  {\bf 3}, 1315 (1987).

\bibitem{resto}
R. M.~Bionta {\it et al.},
Phys.\ Rev.\ Lett.\ {\bf 58} (1987) 1494;
K.~Hirata~{\it et~al.},
Phys.\ Rev.\ Lett.\ {\bf 58} (1987) 1490;
E.N.~Alekseev~{\it et~al.},
Phys.Lett.B {\bf 205} (1988) 209.

\bibitem{ll}
T.~J.~Loredo and D.~Q.~Lamb,
  Phys.\ Rev.\  D {\bf 65}, 063002 (2002);
G.~Pagliaroli~{\it et~al.},
LNGS-TH/08-01 and
  astro-ph 0807.1301.

\bibitem{deruj}
A.~De Rujula,
  Phys.\ Lett.\  B {\bf 193} (1987) 514;
V. Berezinsky~{\it et~al.},
Nuovo Cim. C11 (1988) 287.

\bibitem{imsh}
V.~S.~Imshennik and O.~G.~Ryazhskaya,
  Astron.\ Lett.\  {\bf 30} (2004) 14.


\bibitem{Heger:2002by}
  A.~Heger, C.~L.~Fryer, S.~E.~Woosley, N.~Langer and D.~H.~Hartmann,
  Astrophys.\ J.\  {\bf 591} (2003) 288



\bibitem{Fryer}
  C.~L.~Fryer,
  Nuovo Cim.\  {\bf 121B} (2006) 1233.



\bibitem{Amati:2000ad}
  L.~Amati {\it et al.},
  Science {\bf 290} (2000) 953

\bibitem{Margutti:2007ew}
  R.~Margutti {\it et al.},
  arXiv:0712.1412 [astro-ph].

\bibitem{Woosley:2006fn}
  S.~E.~Woosley and J.~S.~Bloom,
  Ann.\ Rev.\ Astron.\ Astrophys.\  {\bf 44} (2006) 507

\bibitem{Fynbo:2006mw}
  J.~P.~U.~Fynbo {\it et al.},
  Nature {\bf 444} (2006) 1047

\bibitem{Dermer:2007tj}
  C.~D.~Dermer,
  arXiv:astro-ph/0703223.

\bibitem{Jaikumar:2006qx}
  P.~Jaikumar, B.~S.~Meyer, K.~Otsuki and R.~Ouyed,
  arXiv:nucl-th/0610013.




\end{thebibliography}

\end{document}